\documentclass[a4paper,12pt]{article}
\usepackage{graphicx}

\def\beq{\begin{eqnarray}}
\def\eeq{\end{eqnarray}}
\def\bea{\begin{eqnarray}}
\def\eea{\end{eqnarray}}
\begin{document}
\title{Gauge-Higgs unification with brane kinetic terms}
\author{Alfredo Aranda\thanks{fefo@ucol.mx}\\
{\small\it Facultad de Ciencias, Universidad de Colima,}\\
{\small\it Bernal D\'{\i}az del Castillo 340,}\\
{\small\it Colima, Colima, M\'exico}\\
J. Lorenzo D\'iaz-Cruz\thanks{Lorenzo.Diaz@fcfm.buap.mx} \\
{\small\it Cuerpo Acad\'emico de Part\'iculas, Campos y Relatividad}\\
{\small\it FCFM-BUAP, Puebla, Pue. 72570, M\'exico }}

\date{}
\maketitle \vspace{-0.7cm}
\begin{abstract}
\noindent By identifying the Higgs field as an internal component of
a higher dimensional gauge field it is possible to solve the little
hierarchy problem. The construction of a realistic model that
incorporates such a gauge-Higgs unification is an important problem
that demands attention. In fact, several attempts in this direction
have already been put forward. In this letter we single out one such
attempt, a 6D SU(3) extended electroweak theory, 
where it is possible to obtain a Higgs mass
prediction in accord with global fits. One shortcoming of the model
is its prediction for the Weinberg angle, it is too large. We
slightly modify the model by including brane kinetic terms in a way
motivated by the orbifold action on the 6D fields. We show that in
this way it is possible to obtain the correct Weinberg angle while
keeping the desired results in the Higgs sector.
\end{abstract}

\newpage
\section{Introduction}
\label{intro} The little hierarchy problem consists on the
following: treating the standard model as a low-energy effective
theory valid up to a scale $\Lambda$, the Higgs mass suggested by
global fits of electroweak precision data is natural for $\Lambda
\sim 500$~GeV. However, there are bounds at present coming from four
fermion operators that demand $\Lambda \sim 10$~TeV. Thus there is
an order of magnitude discrepancy.

A good amount of work has recently been devoted to find a solution
to this problem. As it is well known, supersymmetry is at the moment
the best candidate for the solution of the hierarchy problem. In
this context, the discrepancy in scales is translated to a
discrepancy between the Higgs mass $m_h$ and the scale of sparticle masses
$m_{SUSY}$. Depending on the specific model, MSSM, NMSSM, etc., there
are several proposals that attempt  to solve or ameliorate the little
hierarchy problem~\cite{Gogoladze:2005az,Choi:2005hd,Dermisek:2005ar,Birkedal:2004xi}.
Other interesting scenario is that of Little Higgs models with and
without T-Parity~\cite{Cheng:2004yc,Cheng:2003ju,Wacker:2002ar}, where
the Higgs boson is identified as a pseudo-goldstone boson of some
unspecified strongly interacting sector.

A different line of work is that of theories in extra dimensions where
the Higgs boson is an internal component of a gauge field of some extended
electroweak symmetry. This idea is not
new~\cite{Manton:1979kb,Fairlie:1979at} and recently has attracted
attention as an
alternative~\cite{Hatanaka:1998yp,Dvali:2001qr,Arkani-Hamed:2001nc,
Antoniadis:2001cv,Csaki:2002ur,Scrucca:2003ra}. In particular,
Scrucca, Serone, Silvestrini and Wulzer~\cite{SSSW04} presented a
complete analysis of an SU(3) electroweak gauge theory in six
dimensions where the two extra dimensions have the geometry
$T^2/{\mathcal Z}_N$. They find that it is possible to formulate a
theory with just one Higgs doublet that gives the prediction (at
leading order): $m_H=2m_W$. At its barest, their model gives a nice
result with little effort. However, in order to get closer to a
realistic theory there are a couple of issues that need to be
addressed. One of them is the stability of the electroweak scale
which has in fact been studied thoroughly for this model
in~\cite{Wulzer:2004ir}, and more generally in~\cite{Biggio:2004kr}
(see also~\cite{Serone:2005ds}). The second issue is the fact that
the prediction obtained for $\tan\theta_W$ is larger than the
correct value.

In this letter we present a simple extension of the original
framework where it is possible to obtain the correct value of the
Weinberg angle while keeping similar results in the Higgs
sector. The basic idea is to introduce brane kinetic terms for the
components of the gauge field in a way motivated by the orbifold
(geometrical) construction of the theory and see that their
inclusion allows us to fix the problem. In Section~\ref{su3} we
review the basic results obtained in~\cite{SSSW04} and stress the
problems mentioned above. Then, in Section~\ref{bkt} we present our
extension based on the inclusion of brane kinetic terms in the
theory. We present an example in detail and comment on other
possibilities to then finish with our conclusions.

\section{SU(3) in 6D}
\label{su3} In this section we review the model presented
in~\cite{SSSW04}. As mentioned in the Introduction, the basic idea
is to relate the extra components of extra-dimensional gauge bosons
to the 4D Higgs field.

Consider an SU(3) gauge theory in six dimensions, two of which are
compactified on a $T^2/{\mathcal Z}_N$ orbifold ($T^2$ is the torus). 
Different choices of $N$ lead to different possibilities for the Higgs
fields; for example for $N=3,4,6$ one can construct models with a single 
Higgs doublet~\cite{SSSW04}. In this letter we are interested in these
models and will consider the case $N=3$.

Gauge bosons are denoted as $A_{\hat{\mu}}$, with the 6D index 
$\hat{\mu}=0,1,2,3,5,6$ split into $\mu = 0, \ 1, \ 2, \ 3$ and $M=5, \ 6$. 
The full gauge symmetry is broken by the orbifold 
boundary conditions (O.B.C.)
in such a way that the gauge symmetry can be broken as: $G \to H$, with 
$H={\rm SU(2)}_W\times {\rm U(1)}_Y$.
Thus, O.B.C. split the group generators into two sets,
$T^A=\{T^a, T^k\}$, where $T^A \, \epsilon \, G$ and $T^a \, \epsilon \,H$.
From this one obtains that $A^a_\mu$ and $A^a_M$ have zero modes in the spectrum
whereas $A^k_\mu$ does not. Note that a vacuum expectation value for
$A^a_M$ can break the symmetry further, namely from $H \to H'=U(1)_{em}$

Then, in the SU(3) model that we are considering, 
the orbifold action on the gauge fields $A_{\hat{\mu}}$,
is such that the invariant components become 
$W_{\mu}=A_{\mu} = \sum_aA_{\mu a}\frac{\lambda^a}{2}$ and
$H_M = A_M = \sum_a A_{M b}\frac{\lambda^b}{2}$, where $a=1, \ 2, \ 3$;
$b=4, \ 5, \ 6, \ 7$ and $\lambda$ are the Gell-Mann matrices. 
Note that these expressions are valid for the 6D gauge
fields and so after compactification we identify the usual 4D gauge bosons with
the zero modes of the Kaluza-Klein tower. From now on we concentrate only on
these zero modes and will omit any ($0$) superscript on the 4D fields.

The zero modes of the 4D vector fields above can be expressed in matrix notation as
\beq \label{matrixbosons} \nonumber
A_{\mu}&=&\frac{1}{2}\left(\begin{array}{ccc}
  A_{\mu}^{(3)}+\frac{1}{\sqrt{3}}A_{\mu}^{(8)} & A_{\mu}^{(1)}-iA_{\mu}^{(2)} & 0\\
  A_{\mu}^{(1)}+iA_{\mu}^{(2)} & -A_{\mu}^{(3)}+\frac{1}{\sqrt{3}}A_{\mu}^{(8)} & 0 \\
  0 & 0 & -\frac{2}{\sqrt{3}}A_{\mu}^{(8)} \end{array}\right) \\
&=&
\frac{1}{2}\left(\begin{array}{ccc}
  W_{\mu}^{(3)}+\frac{1}{\sqrt{3}}B_{\mu}^{(8)} & \sqrt{2}W_{\mu}^+ & 0\\
  \sqrt{2}W_{\mu}^- & -W_{\mu}^{(3)}+\frac{1}{\sqrt{3}}B_{\mu}^{(8)} & 0 \\
  0 & 0 & -\frac{2}{\sqrt{3}}B_{\mu}^{(8)} \end{array}\right) \, ,
\eeq
where we have introduced the usual SU(2) notation. In turn the 4D
scalar fields are identified as
\begin{eqnarray}\label{matrixhiggs} \nonumber
  H_M &=& \frac{1}{2} \left( \begin{array}{ccc}
    0 & 0 & A_M^{(4)}+iA_M^{(5)} \\
    0 & 0 & A_M^{(6)}+iA_M^{(7)} \\
    A_M^{(4)}-iA_M^{(5)} & A_M^{(6)}-iA_M^{(7)} & 0 \end{array}\right) \\
  &=&
  \frac{1}{\sqrt{2}} \left( \begin{array}{ccc}
    0 & 0 & H^{* +}_M \\
    0 & 0 & H^0_M \\
    H^-_M & H^{0*}_M & 0 \end{array}\right) \, .
\end{eqnarray}

Substituting these expressions into the Lagrangian for the zero modes obtained after
integration over the internal torus one gets
\beq
\label{4dlagrangian}
\nonumber
    {\cal L}_4 & = &
    -\frac{1}{4}(F_{\mu\nu}^a)^2 -\frac{1}{4}(B_{\mu\nu})^2
    +\left|\left(
    \partial_{\mu}-ig_4W_{\mu}^a\frac{\tau^a}{2}
    -ig_4\sqrt{3}\frac{1}{2}B_{\mu} \right){\cal H}\right|^2 \\
    &-&\frac{g_4^2}{2}|{\cal H}|^4 \, ,
\eeq with $g_4 = g_6/(\frac{2}{N}\pi\sqrt{R_5R_6})$ as the 4D gauge
coupling and ${\cal H}^T=\left(\begin{array}{cc}
  H^{* +}_M & H^0_M \end{array}\right)$
the Higgs doublet. $R_{5,6}$ denote the radii of the torus.

One can immediately obtain some interesting Higgs physics results
out of this simple model. Note that a Higgs tree-level potential is
present~\footnote{Thus the model
also realizes the unification of quartic Higgs and gauge
couplings \cite{ourXDGHU}, without Supersymmetry.}: 
this is to be contrasted with the 5D case where this is not
the case and one obtains Higgs masses that are very
small~\cite{SSSW04}. In the present case, if one makes the (strong)
assumption that local operators (tadpole operators) have no
significant effect on the potential and ignoring logarithmic
divergences, the leading one-loop effective potential for the Higgs
is \beq\label{potential} V({\cal H})=-\mu^2|{\cal
H}|^2+\lambda|{\cal H}|^4 \, , \eeq where $\mu^2$ is generated
radiatively. By assuming that $\mu^2>0$ and setting $\langle|{\cal
H}|\rangle=v/\sqrt{2}$ one obtains
$m_H=\sqrt{2}\mu=\sqrt{2\lambda}v$ and $m_W=gv/2$. Taking the ratio
leads to

\beq\label{ratio}
\frac{m_H}{m_W}=\frac{2\sqrt{2\lambda}}{g_4}=2 \, ,
\eeq
where we use the fact that $\lambda = g_4^2/2$. 

It is interesting that this simple model leads to a Higgs mass in
approximately the right range suggested by global fits. In order to
make it more realistic however one needs to understand the details
regarding the tadpole operators that give the main radiative
corrections to the result. As mentioned in Section~\ref{intro}, the
origin of these operators has been studied in~\cite{Biggio:2004kr}
where they found that tadpoles are always allowed in orbifolds
compactifications based on $T^{D-4}/{\mathcal Z}_N$ ($D$ even,
$N>2$). Another interesting result is that on $T^{D-4}/{\mathcal
Z}_2$ with arbitrary $D$, tadpoles can only appear in $D=6$ (except
for models with only bulk gauge fields). In~\cite{SSSW04}, an
argument was presented that a globally vanishing one-loop tadpole is
indeed harmless for the stability of electroweak symmetry breaking.
Thus, constructing a model with globally vanishing tadpoles is a way
to go and one can do this by introducing a suitable fermion
content~\cite{Wulzer:2004ir}.

Another problem with the model is that it gives
$\tan\theta_W=\sqrt{3}$, which is larger than the measured value. It
is therefore interesting to see if one can find modifications of
this model that would fix this problem and at the same time keep all
the nice features in the Higgs sector. As suggested
in~\cite{SSSW04}, it might be possible to lower the value of
$\tan\theta_W$ by adding extra U(1) symmetries, however there can be
other possibilities. In this letter we present an alternative where
the basic idea is to introduce brane kinetic terms in the 6D theory
and explore their effects on the Higgs mass.

\section{Brane kinetic terms}
\label{bkt}
In this section we describe how the addition of localized brane kinetic
terms can be used to lower the prediction for $\tan\theta_W$. A discussion on brane kinetic
terms and their physical implications can be found in~\cite{APS03,CPTW03,DHLR04,Chaichian:2002uy}.
As mentioned in the previous section, we are interested in models with one Higgs doublet
an will present our analysis for an SU(3) theory compactified on $T^{2}/{\mathcal Z}_3$.

\subsection{Brane kinetic terms at a point}
We start with the following 6D Gauge Lagrangian:
 \beq \label{modela} {\cal
L}_{6D}=-\frac{1}{4}\left(\left(F_{\hat{\mu}\hat{\nu}}\right)^2
+\delta(x_5)\delta(x_6)\left[c_1\left(F^{(a)}_{\mu\nu}\right)^2
+c_2\left(F^{(8)}_{\mu\nu}\right)^2\right]\right) \, , \eeq where
again $a = 1, 2, 3$. In eq.~(\ref{modela}) we have added a localized
kinetic term to the SU(3) 6D gauge theory at the fixed point
$x_5=x_6=0$~\footnote{We can add such terms in every fixed point. In
this case we concentrate on one for clarity.} and taken $c_1$ and
$c_2$ as positive constants with mass dimension $-2$. Note that we
have introduced two different strengths in the localization terms.
This choice is certainly a source of fine tuning in the model and at this 
moment we do not have a precise argument for it but to say
that it is motivated by the geometrical breaking of the symmetry,
where the orbifold action is already differentiating the
components of the gauge fields. We will see that this
differentiation will play a crucial role in determining a correct value
for $\tan\theta_W$. In this particular example and for simplicity, we
have added brane kinetic terms only to those components of the 6D
gauge fields that will become 4D gauge fields and not to the scalar
ones.

After compactification the 4D Lagrangian for the zero modes becomes (in SU(2) notation)
\beq\label{l4dsu2}\nonumber
      {\cal L}_{4D}^{(0)} & = &
      -\frac{1}{4}(F_{\mu\nu}^a)^2 -\frac{1}{4}(B_{\mu\nu})^2
      +\left|\left(
      \partial_{\mu}-i\frac{g_4}{\sqrt{{\cal Z}_1}}W_{\mu}^a\frac{\tau^a}{2}
      -i\frac{g_4}{\sqrt{{\cal Z}_2}}\sqrt{3}\frac{1}{2}B_{\mu} \right){\cal H}\right|^2 \\
      &-&\frac{g_4^2}{2}|{\cal H}|^4 \, ,
\eeq
where
\beq
    {\cal Z}_{1,2}=1+\frac{c_{1,2}}{(2\pi/3)^2 R_5 R_6} \ .
\eeq
These factors, ${\cal Z}_{1}$ and ${\cal Z}_{2}$, have been introduced
in order to properly normalize the fields in the 4D effective theory.

As before, we obtain a tree-level quartic Higgs coupling. 
Note however that it is now possible
to fix the correct value for $\tan\theta_w$ by a suitable choice of
${\cal Z}_{1,2}$. This is where the effect of having different strengths for the brane kinetic
terms appears. In order to see this explicitly we define $g=g_4/\sqrt{{\cal Z}_1}$
and $g^{\prime}=\sqrt{3}g_4/\sqrt{{\cal Z}_2}$ as the SU(2)$_W$ and U(1)$_Y$
gauge couplings respectively. Then, using the same arguments that led to eq.~(\ref{ratio}),
we obtain

\beq
\tan\theta_W=\frac{g^{\prime}}{g}=\sqrt{\frac{3{\cal Z}_1}{{\cal Z}_2}} \,
\eeq
while the Higgs boson mass is now given by
\beq
\frac{m_H}{m_W}=2\sqrt{{\cal Z}_1} \ .
\eeq

In order to explore some of the parameter space, we consider the particular case of a
torus of dimensions $R\equiv R_5=R_6$. We then identify a compactification scale
$M_C=1/R$ for each of the two extra dimensions.

\begin{figure}[ht]
  \begin{center}
    \includegraphics[width=10cm]{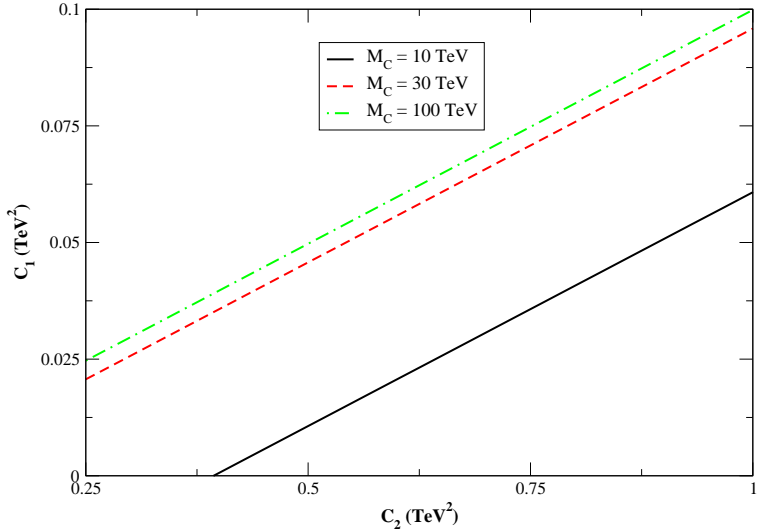}
    \caption{Values of $c_1$ and $c_2$ consistent with the correct value of $\tan\theta_W$ for
different choices of the compactification scale $M_C$. We use $\tan\theta_W=0.54839$ in these
results~\cite{PDG}.}
    \label{fig:c1c2}
  \end{center}
\end{figure}

Figure~\ref{fig:c1c2} shows the values of $c_1$ and $c_2$ consistent
with the correct value of $\tan\theta_W$ for three choices of
$M_C=1/R$: $10$, $30$ and $100$~TeV. We see that in all three cases,
a solution requires $c_2>c_1$. Furthermore, solutions exist for
$c_2\leq 1$. This is relevant since our analysis incorporates only
the zero modes and thus is valid only for the case in which the both
$c_1$ and $c_2$ are small. Larger values would require a systematic
study of KK modes mixing and its implication on the stability of the
electroweak sector. We are performing a study to quantify this
effect due to KK mixing for a general class of models of this type~\cite{ADfuture}.

\begin{figure}[ht]
  \begin{center}
    \includegraphics[width=10cm]{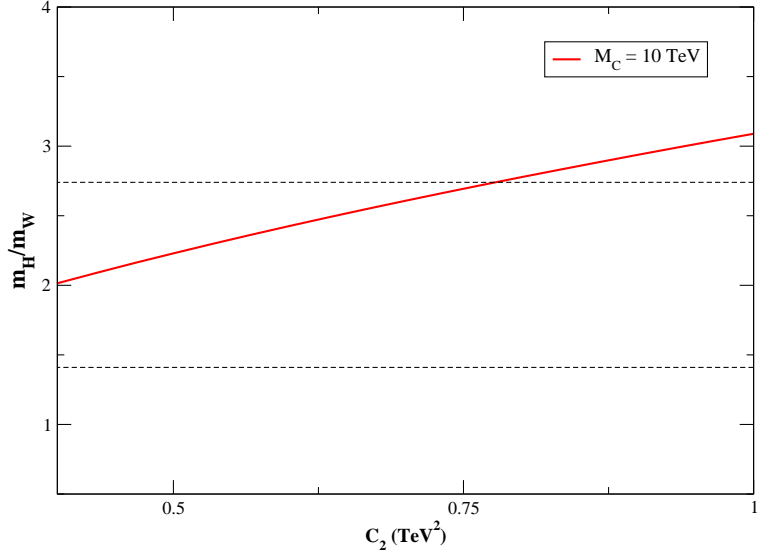}
    \caption{$m_H/m_w$ obtained with the values of $c_1$ and $c_2$ in fig.~\ref{fig:c1c2} for
      $M_C=10$~TeV. The horizontal lines correspond to the range consistent with global fits.}
    \label{fig:ratio}
  \end{center}
\end{figure}

Using these results, we present the prediction for the ratio
$m_H/m_W$ in figure~\ref{fig:ratio}. The result is plotted as a
function of $c_2$ where for each $c_2$ we have used the value of
$c_1$ presented in fig.~\ref{fig:c1c2}. We also plot the range
suggested for this ratio by global fits (horizontal lines). Note
that the allowed parameter space is consistent with
the conditions described above for the case $M_C=10$~TeV. 
In the case of larger $M_C$ ($30$ and
$100$~TeV) the ratio is off the scale for all the values of $c_2$ and $c_1$
consistent with $\tan\theta_W$.

We stress that while the Higgs mass can be in the range suggested by
EW precision data, there are large regions of  parameter space 
above $2m_W, 2m_Z$  which are certainly permitted. Therefore, a detailed study
of Higgs decays $H\to WW$ and $H\to ZZ$ at future facilities (LHC, ILC)
will help to put strong constraints on this type of models \cite{ourhVV}. 

This example shows that it is possible to fix the problem in the
original model by introducing brane kinetic terms as in~(\ref{modela}). As discussed
above, we introduced brane kinetic terms only for the 6D gauge components that turn
into 4D gauge components. Apart from simplicity, our goal was to relate these terms to
the value of $\tan\theta_W$ without disturbing the original tree level scalar potential.
However, brane kinetic terms for the gauge-scalar components can also be incorporated 
and will cause modifications to the classical scalar potential~\cite{ADfuture}. 

\section{Conclusion}\label{conclusion} \vspace{12pt}

In order to solve the little hierarchy problem it is possible to
identify the Higgs field with components of gauge fields in higher
dimensional electroweak theories. One such extended electroweak
theory was presented by Scrucca {\it et. al.} in~\cite{SSSW04},
where an SU(3) gauge theory in six dimensions is acted upon by an
orbifold in such a way that one obtains a Higgs doublet in the
low-energy effective theory. The model predicts $m_H=2 m_W$, which
is in the range suggested by global fits, and $\tan\theta_W =
\sqrt{3}$ which is larger than the measured value. In this letter we
presented a modification of the model in~\cite{SSSW04} that fixed
the prediction of the Weinberg angle while keeping the results in
the Higgs sector. We accomplished this by introducing brane kinetic
terms in the 6D theory with different strengths for the already
(orbifold) differentiated components of the gauge fields.

\section*{Acknowledgments} A.A. would like to thank Paolo Amore
for reading the manuscript and also acknowledges support from Conacyt grant
no. 44950 and PROMEP.

\end{document}